# Enhancing thermoelectric performance of 2D Janus ISbTe by strain engineering: A first principle study

Anuja Kumari,[1, a] Abhinav Nag,[2] Santanu K. Maiti,[1, b] Jagdish Kumar[3, 4]

[1]*Physics and Applied Mathematics Unit, Indian Statistical Institute, 203 Barrackpore Trunk Road, Kolkata-700 108, India*
[2]*Department of Physics, GGDSD college, Rajpur, Palampur-176061, India*
[3]*Department of Computational Sciences, Central University of Punjab, Bhatinda-151 401, India*
[4]*Department of Physics, Central University of Himachal Pradesh, Dharamshala-176 215, India*
[a]anujakumari_v@isical.ac.in
[b]santanu.maiti@isical.ac.in

Recent developments in the 2D materials laid emphasis on finding the materials with robust properties for variety of applications including the energy harvesting. The recent discovery of Janus monolayers with broken symmetry has opened up new options for engineering the properties of 2D layered materials. Present study focuses on enhancing thermoelectric properties of 2H-ISbTe 2D Janus monolayer. All the calculations have been performed using fully relaxed unit cell and employing the pseudo potential based quantum espresso code. Calculated structural parameters are in good agreement with previous literature reports. The lattice dynamics calculations predicts this monolayer can withstand a strain of up to 4% beyond which imaginary frequencies appear in the phonon dispersion curves. Computed electronic structure reveals that the monolayer is an indirect wide bandgap material and the bandgap decreases with tensile strain. Furthermore, the computed thermoelectric properties show that the studied monolayer has high Seebeck coefficient of ~ 300 μV/K and low thermal conductivity which yields reasonably high ZT of ~ 1.31 for a strain of 2% at 300 K with p-type doping. Therefore, our study signifies the fact that tensile strain and p-type doping of 2D Janus monolayer ISbTe can enhance ZT from 0.87 to 1.31 at room temperature which makes it a promising candidate for thermoelectric applications.

## 1. Introduction

The prime concern nowadays is scavenging energy to meet the abrupt increase in global energy demands [1]-[4]. Various technologies were developed by scientists and thermoelectricity is one such area that has huge scope for investigation due to the abundance of waste heat [5]-[11]. To convert waste heat into useful electrical energy by using thermoelectric generators is known for decades for their deep space applications etc [12]-[17]. Efficiency of any thermoelectric generator is a matter of investigation and needs attention [18]-[20]. Thermoelectric generators (TEGs) mainly consist of highly doped p-type and n-type semiconductors that are connected in parallel thermal and series electrical connections. For an efficient TEG module, these p-type and n-type semiconductors should possess characteristics as phonon glass and electron crystal (PGEC) [20]-[22]. To improve efficiency of TEGs, many material scientists have explored different TE materials with good thermoelectric properties but achieving high efficiency is yet a challenge [19],[20]. For any given material efficiency is measured by dimensionless figure of merit (FOM) represented by ZT as [22],

$$ZT = \frac{S^2 \sigma T}{K}. \qquad 1.1$$

In the above equation, S represents the Seebeck coefficient, σ represents the electrical conductivity, and *K* represents thermal conductivity (sum of electronic thermal conductivity and phononic thermal conductivity). Here, factor $S^2\sigma$, indicates the electric performance of the material also known as power factor (PF). For high efficiency, PF should be high and K should be low at a given temperature [23]-[25]. However, for high PF, Seebeck coefficient (S) and electrical conductivity (σ) should be high but simultaneously electronic part of thermal conductivity ($K_e$) also increases which result in low value of ZT. Thus, optimizing these coupled parameters simultaneously to obtain high ZT is a challenging task [26]-[28]. Various strategies to optimize these parameters such as power factor augmentation by carrier filtering [29]-[33], carrier pocket engineering [34]-[36], complex structures [37]-[40], low dimensional structures [41]-[44] and valley degeneracy [45]-[49] etc., that improves ZT significantly were reported in literature.

Due to robust transport properties of graphene and layered materials, wide scope for exploration among 2D layered materials has opened up for thermoelectric applications [50]-[53]. Also, their significant structural, physical, chemical properties and feasible fabrication methods using chemical vapor deposition (CVD), physical vapor deposition, mechanical exfoliation etc., make them potential TE materials for thermoelectric applications [54]-[56]. In a work Gu and Yang [57] suggested lowering the lattice thermal conductivity by changing the stoichiometry of the structure. Using this mechanism, recently the synthesis of MoSSe monolayer [58] by selenisation of $MoS_2$ discovered new family of 2D materials named Janus materials. These 2D monolayers have exhibit out of plane symmetry that results in significant properties. They possess an extra degree of freedom to tune material properties while retaining the exotic characteristics of their parent structure. Numerous work in the literature suggested that these Janus materials can be a potential material for TE applications [59]-[62]. Many Janus monolayer of transition metal dichalcogenides (TMDCs) were reported with high ZT, such as Janus monolayer WSTe at higher temperature shows a high ZT of 2.56 [63], another study by S. Zhao *et al.* [64] for Janus monolayer ZrSSe with 6% strain have reported high ZT of 4.41 and 4.88 for p-type and n-type materials. First principle study reveals intrinsic magnetism and asymmetry in some other Janus monolayers such as manganese chalcogenides [65], chromium trihalides [66] and lanthanum bromiodide [67] to be magnetic Janus monolayers.

Due to layered structures these Janus monolayers have lower thermal conductivity as compared to their parent structures and further strategies were employed by the researchers to improve its ZT. Recently a study by Chu *et al*. [68] predicts that Janus monolayers of ISbTe have lower thermal conductivity ~ 1.5 Wm/K in its 2H phase as compared to 1T phase and another study by Guo *et al*. predict ATeI (A=Sb and Bi) to be a potential thermoelectric material with ZT ~1.11 in 1T phase. Also, a study by Kumari *et al*. [69] reports that biaxial tensile strain on Janus monolayers of TiXY (X,Y=S,Se,Te) enhances ZT ~1.9. Motivated by these studies, *we have investigated effect of biaxial tensile strain on the structural, electronic and thermoelectric properties of ISbTe Janus monolayer in 2H phase which have lower thermal conductivity*. Janus monolayer of ISbTe in 2H phase crystallizes in the trigonal space group P3m1 (No. 156). Figure 1 depicts the top view and side view of this monolayer. In 2H phase this monolayer consists of three-atomic thick layer with Sb atomic layer sandwiched between two different halogen atomic layers, I and Te, keeping reflection asymmetry on the mirror plane.

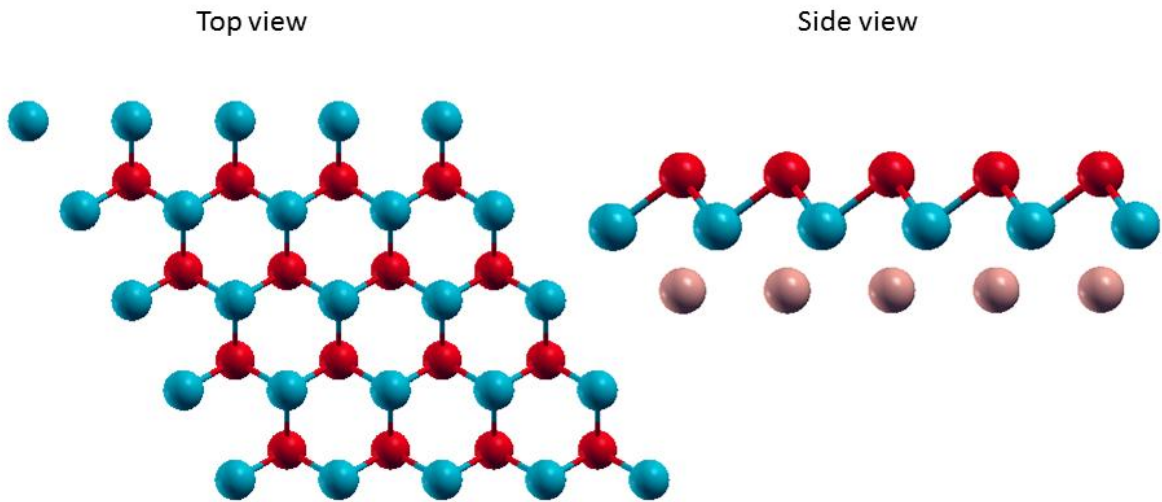

Figure 1: Crystal structure of 2D Janus monolayer of ISbTe (P3m1: no. 156) top view and side view with I atoms in pink, Sb atoms in blue, Te atoms in red color.

Present study primarily focuses on the effect of biaxial tensile strain on the dynamical stability, structural properties, electronic properties, and thermoelectric properties of 2H ISbTe Janus monolayer under strain. Our study signifies that strain can be an effective strategy to improve figure of merit of this material. The main findings of our work are i) phonon dispersion reveals that material is dynamically stable up to 4% biaxial tensile strain, ii) calculated band structure implies that with increase in strain band gap increases, iii) at 300K we get high Seebeck coefficient and lower thermal conductivity with p-type doping, iv) computed figure of merit (FOM) of unstrained system enhances from 0.87 to 1.31 with 2% strain at room temperature, v) thermoelectric properties can be further tuned with strain and chemical doping at a given temperature. Thus, present study supports that this material can be a potential material for thermoelectric applications even in the low temperature regime.

The factual arrangement in present paper is described as follows. Theoretical formalism and detailed computational parameters were described in Sec. 2. In Sec. 3 we discuss detailed outcomes as computed lattice parameters, phonon dispersion curves and thermoelectric coefficients. Lastly conclusion of this study is given in Sec. 4.

## 2. Computational Details and Theoretical Framework

Present computational study was performed using density functional theory (DFT) based formalism as implemented in pseudo-potential-based Quantum espresso code [70]. Janus monolayer of ISbTe is fully relaxed to obtain optimized lattice parameter. Further, optimized crystal structures were utilized to compute the phonon dispersion curves using *density functional perturbation theory* (DFPT) formalism in quantum espresso. For considering the exchange and correlation interaction among the electrons the *generalized gradient approximation* within the framework of Perdew-Burke-Ernzerhof (PBE) formalism is used [71]. All the computations were done using energy cutoff of 350 eV with cold smearing. For considering the Brillouin zone (BZ) sampling, the well converged Monkhorst-Pack of 12×12×1 *k*-points for calculating electronic properties and 36×36×2 for calculating thermoelectric properties are used. Structural optimization was achieved by relaxing the crystal structures to a force tolerance of $10^{-4}$ eV/atom. To avoid the interaction among the vertical layers along z-axis a large vacuum space of 15 Å is considered. Further, phonon dispersion curves were computed using q-point set of 2×2×1 with tetrahedron method. Once the electronic band structures were obtained, using the semi-classical Boltzmann transport formalism as implemented in the BoltzTrap code [72] transport properties were computed within constant scattering time approximation (CSTA). According to CSTA relaxation time is assumed as independent of energy and finally ZT have no dependence on relaxation time. This formalism has been proved to significantly report the electronic part of ZT. According to the semi-classical Boltzman theory, *energy projected conductivity tensor '$\Theta(\varepsilon)$'* is related to conductivity tensor by the relation

$$\Theta(\varepsilon) = \frac{1}{N} \sum_{k} \sigma(k) \left( -\frac{\delta(\varepsilon - \varepsilon_k)}{d\varepsilon} \right) \ . \qquad 2.1$$

Here, $N$ represents the total number of k-points in BZ, $k$ signifies band index and $\varepsilon$ stands for specific energy of electrons. For a given crystal with unit cell volume $\Omega$, integrating this energy projected conductivity tensor we can evaluate the transport coefficients such as electrical conductivity, electronic part of thermal conductivity and Seebeck coefficient as

$$\sigma(T,\mu) = \frac{1}{\Omega} \int \Theta(\varepsilon) \left[ -\frac{\delta f_\mu(T,\varepsilon)}{\delta k} \right] d\varepsilon \ , \qquad 2.2$$

$$\kappa_e(T,\mu) = \frac{1}{eT\Omega} \int \Theta(\varepsilon)(\varepsilon - \mu)^2 \left[ \frac{\delta f_\mu(T,\varepsilon)}{\delta k} \right] d\varepsilon \ , \qquad 2.3$$

$$S(T,\mu) = \frac{1}{eT\Omega T(T,\mu)} \int \Theta(\varepsilon)(\varepsilon - \mu) \left[ \frac{\delta f_\mu(T,\varepsilon)}{\delta k} \right] d\varepsilon \ . \qquad 2.4$$

In the above equations $T$ stands for the equilibrium temperature, $\mu$ represents the chemical potential, $\kappa_e$ is the electronic part of the thermal conductivity and S refers to the Seebeck coefficient. Once we obtain all the thermoelectric coefficients we can compute dimensionless figure of merit ZT by the relation,

$$ZT = \frac{S^2 \sigma T}{\kappa_e + \kappa_l}. \qquad 2.5$$

From equation 2.5 we clearly observe that for more accurate prediction of ZT, phonon contribution $K_l$ and electron contribution $K_e$ towards thermal conductivity are necessary. A study reported by Chu *et al.*[68] predicts low thermal conductivity of ~1.5 W/mK for Janus monolayer

ISbTe in 2H phase. Thus, in these monolayers effect of phononic thermal conductivity is negligibly small and we have ignored phononic part in calculating ZT, without loss of any generality.

## 3. Numerical Results and Discussion

### 3.1 *Structural properties*

Figure 1 represents the supercell of Janus monolayer that has been considered in the present study. Fully optimized crystal structure with a force tolerance of 0.001eV/atom was utilized to obtain the equilibrium lattice parameters for all the compositions. The calculated lattice constant for parent ISbTe within GGA-PBE scheme is 4.13 Å. Janus monolayer of ISbTe have thickness of 4.05 Å with corresponding bond distances are 3.26 Å, 3.00 Å for Sb-I and Sb-Te respectively. Obtained lattice parameters for ISbTe are in good agreement with the literature reports [68]. Further, these structural parameters significantly change with the tensile strain and obtained data is listed in Table 1.

Table 1: Lattice constants and bandgap of various strains on Janus ISbTe systems obtained using GGA-PBE exchange and correlation for electrons.

| S. No. | Tensile strain | Parameter | ISbTe (2H) |
|---|---|---|---|
| 1 | 0% | Lattice constant (Å) | 4.13 (4.19)[68] |
| | | Bandgap (eV) | 1.04 (1.21) |
| 2 | 1% | Lattice constant (Å) | 4.17 |
| | | Bandgap (eV) | 1.14 |
| 3 | 2% | Lattice constant (Å) | 4.21 |
| | | Bandgap (eV) | 1.23 |

### 3.2 Lattice Dynamics and Dynamical Stability

Janus monolayers of ISbTe reported in the recent studies suggest that this material with low young's modulus may undergo readily under strain [68]. Also, strain (ε) has drastic impact on the performance of any thermoelectric material. In this work, we have examined effect of strain on the lattice dynamics of ISbTe by considering systematic strain percentage as $\varepsilon\% = \left(\frac{a-a_0}{a_0}\right) \times 100$ where $a_0$ represents the relaxed lattice constant. We have obtained phonon dispersion curves with various strains for ISbTe Janus monolayers along different high-symmetry directions at the steps of Δε=2% as shown in Figure 2.

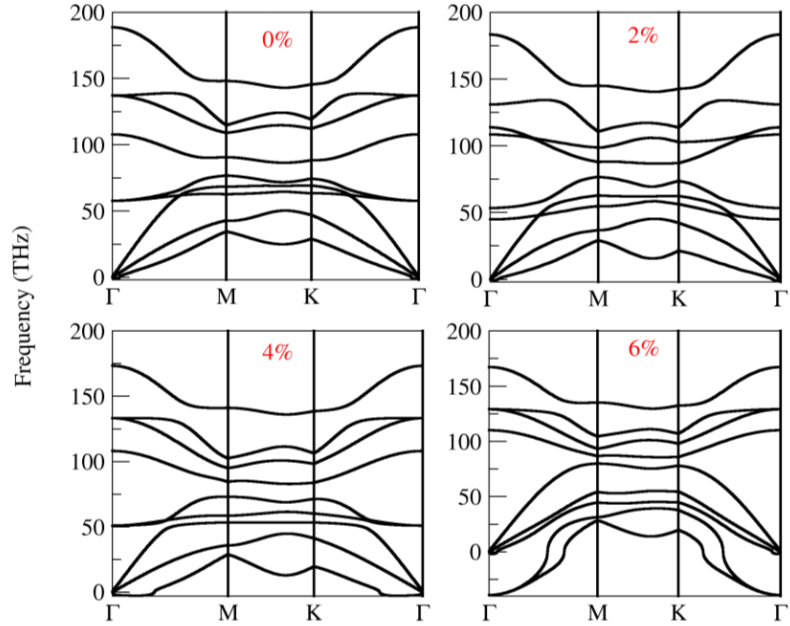

Figure 2: Phonon dispersion for ISbTe at different biaxial tensile strain with Δε=2% along high symmetry k-points

With the first glance of the phonon dispersion curves we observed that dynamically ISbTe is stable up to 4% strain. The computed phonon dispersion along a high symmetry path has three acoustic and six optical branches that signifies in a unit cell three atoms Sb, I, and Te are present. To predict dynamical stability of material phonon frequencies were analyzed for imaginary frequency if any. Dynamically, presence of an imaginary frequency shows the non-restorative force among the atoms and thus decreases in potential energy which signifies instability of the material. Figure 2 represents computed phonon dispersion curves for ISbTe and we observed that there are no imaginary phonon frequencies up to 4% strain. With the further increase in strain to 6% we clearly observe negative phonon frequency. Thus, dynamically ISbTe Janus monolayer can with stand strain up to 4% and beyond this limit application of strain gives rise to imaginary frequencies in the phonon dispersion curve and makes this material unstable.

### 3.3 Electronic Properties

The computed electronic band structures as shown in Figure 3 predict ISbTe to be an indirect band gap semiconductor with 1.04eV band gap with conduction band minima (CBM) at K and valence band maxima (VBM) in between Γ-K high symmetry k-points which is comparable to the literature report [68] as 1.21 eV. Figure 3 illustrates effect of strain on the electronic band structure and we observed that band gap increases to 1.21 eV with 2% strain.

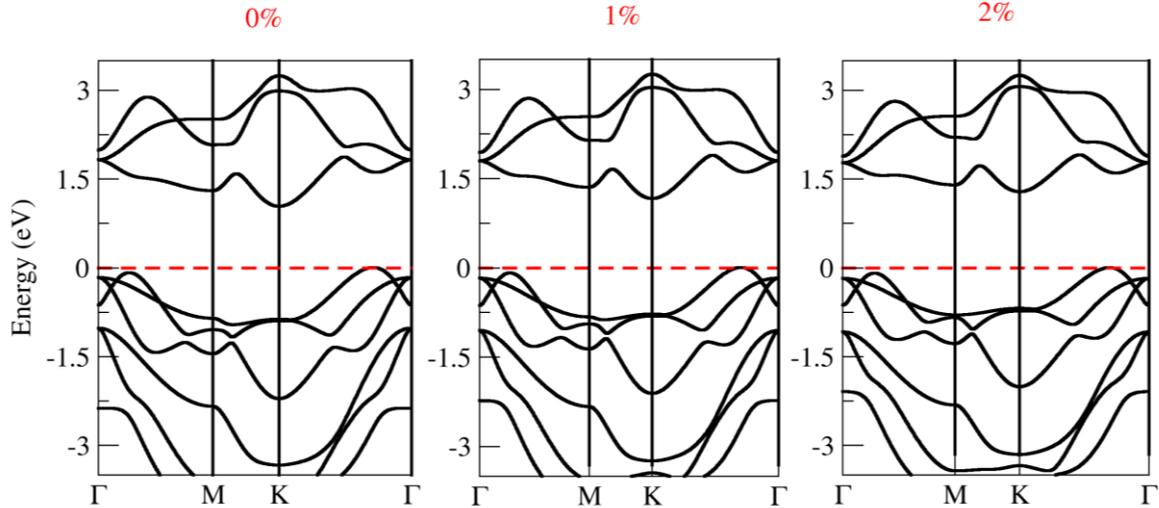

Figure 3: Electronic band structure for various strains on 2H- ISbTe.

The upward shifting of the conduction band maxima (CBM) with applied tensile strain is clearly visible in Figure 3. This significant change can be expected due to application of tensile strain lead to bond stretching which results in significant band rearrangement hence affecting the conduction band minima. The transport properties have strong dependence on the band structure and hence the band rearrangement is expected to significantly affect thermoelectric properties.

**3.3 Thermoelectric Properties**

The computed electronic thermoelectric coefficients using the semi-classical Boltzmann transport theory within constant time approximation with no dependence of relaxation time on thermoelectric coefficients were discussed. We have obtained the Seebeck coefficient (*S*), electrical conductivity (σ), electronic part of thermal conductivity ($\kappa_e$) and power factor $S^2\sigma$ as a function of chemical potential (*μ*) to study the effect of doping and strain effects on the transport properties. The location of chemical potential defines the negative and positive doping levels with electron (*n*-type) and hole (*p*-type) dopings, respectively.

*3.3.1 Seebeck coefficient (S)*

The Seebeck coefficient of a system is the induced electric potential due to thermal gradient. Any good thermoelectric material should possess higher Seebeck coefficient to produce higher electric potential for a given thermal gradient across the material. For a given temperature say *T*, electronic charge *e* and effective mass *m\**, Seebeck coefficient can be defined at a carrier concentration '*n*' as [73],[74]

$$S = \frac{8\pi^2 k_B^2 T}{3eh^2} m^* \left(\frac{\pi}{3n}\right)^{\frac{2}{3}} \qquad 3.1$$

where, $k_B$ is Boltzmann's constant, and *h* is Planck constant. Utilizing the group velocities from the computed electronic band structure within constant scattering time approximation (CSTA) the electronic conductivity tensor $\Xi(\varepsilon)$ can be obtained as discussed in Sec. 2. Further, to study the effect of doping on Seebeck coefficient at different chemical potential, $\mu = \pm 1\text{eV}$ is used. The obtained Seebeck coefficient for the monolayer of ISbTe as a function of carrier concentration is presented in Figure 4. For n-type doping, unstrained monolayer have the highest Seebeck coefficient of ~2890 µV/K at 100K which further decreases to ~2800 µV/K. At, 300K temperature we observe increase in the Seebeck coefficient from ~1545 µV/K to ~ 2153 µV/K with 0% and 2% strain respectively. Similar trend can be seen in the p-type doping with highest Seebeck coefficient ~- 1402 µV/K with 0% strain to ~-1989 µV/K with 2% strain at 300K. Although at 100K and 200K, Seebeck coefficient is subsequently higher than 2500 µV/K.

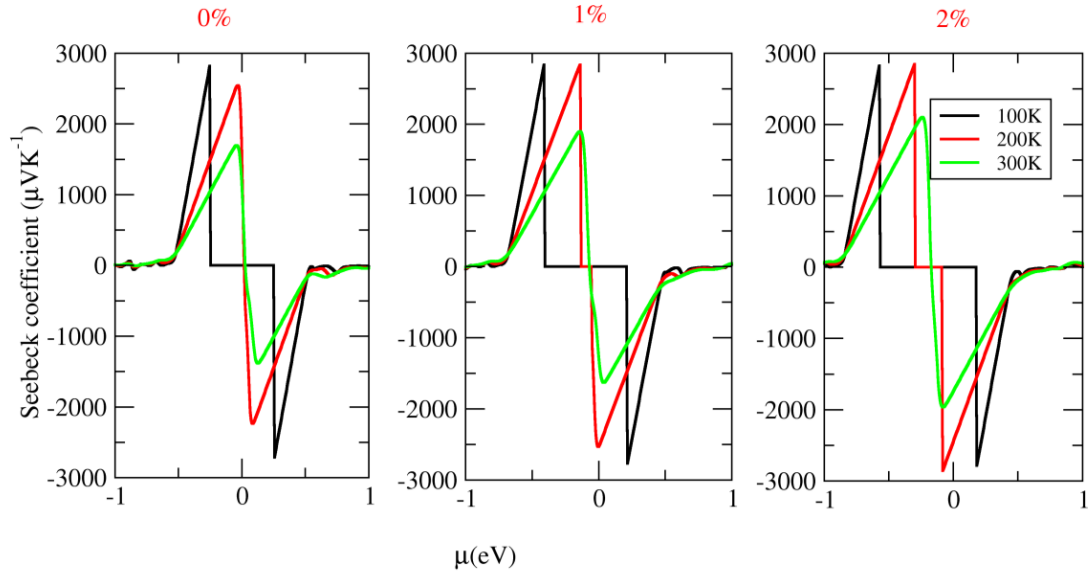

Figure 4: Seebeck coefficient as a function of chemical potential at 100K, 200K, and 300K.

*3.3.3 Thermal conductivity (K)*

To obtain high figure of merit, lower thermal conductivity is desirable. However, it strongly depends upon the electronic and phononic contributions of thermal conductivity. The Wiedemann-Franz law relates these two physical quantities by the relation $\frac{\kappa_e}{\sigma} = LT$ [75], where $\kappa_e$, σ, $L$, $T$ represent the electronic thermal conductivity, electrical conductivity, Lorenz number and absolute temperature respectively. According to this law, for a simplest free electron model, the ratio of the electrical and thermal conductivities is directly related to the product of constant Lorenz umber and absolute temperature. However, in practical applications beyond free electron model, these parameters can be optimized to enhance thermoelectric efficiency.

Figure 5 represents the electronic thermal conductivity as a function of chemical potential at 100K, 200K and 300K. It can be observed from the plots that for n-type doping, an unstrained system have lower thermal conductivity of ~ $1.9 \times 10^{14} W/mKs$ at 100K which increases to $5.4 \times 10^{14} W/mKs$ at 300K. With further 1% increase in strain led to decrease in thermal conductivity to $1.1 \times 10^{14} W/mKs$ at 100K to $3.8 \times 10^{14} W/mKs$ at 300K. Further increase in strain drecreases the thermal conductivity to $0.9 \times 10^{14} W/mKs$ at 100 K to $2.1 \times 10^{14} W/mKs$ at 300K. Similarly, for p-type doping unstrained monolayer have thermal conductivity $4.5 \times 10^{14} W/mKs$ at 300K and further increase in strain increases the thermal conductivity $> 4.5 \times 10^{14} W/mKs$ at 300K. Thus, observed values of thermal conductivity suggest that n-type doping is favourable for obtaining lower thermal conductivity at 300K.

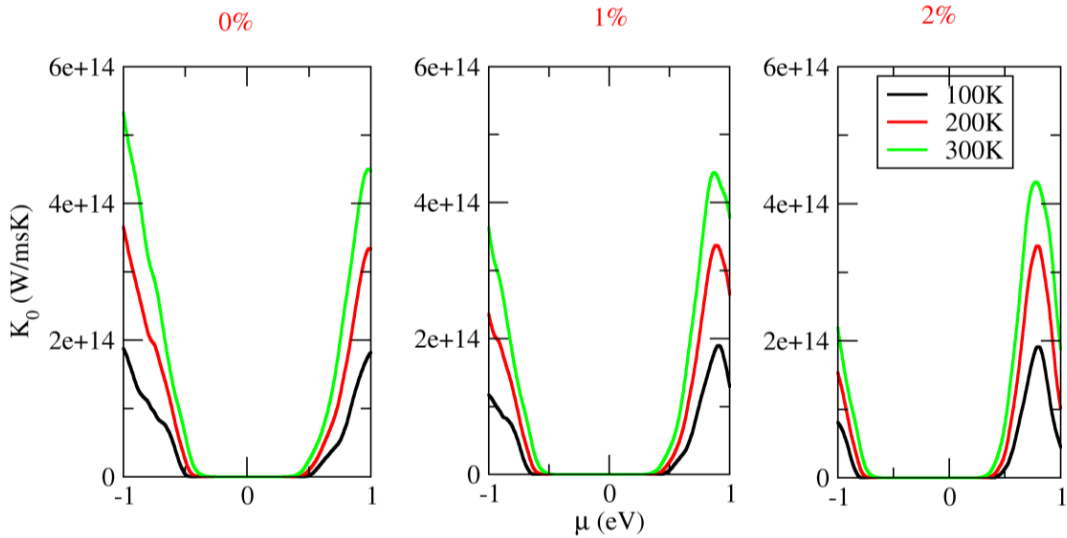

Figure 5: Variation of electronic part of thermal conductivity with chemical potential at various temperature regions.

*3.3.4 Power factor (S$^2$σ)*

The electronic performance of a material can be defined by a parameter known as power factor as PF= $S^2\sigma$. Here, Seebeck coefficient and electrical conductivity are taken into account. Thus to obtain high figure of merit ZT for given TE material high PF is essential.

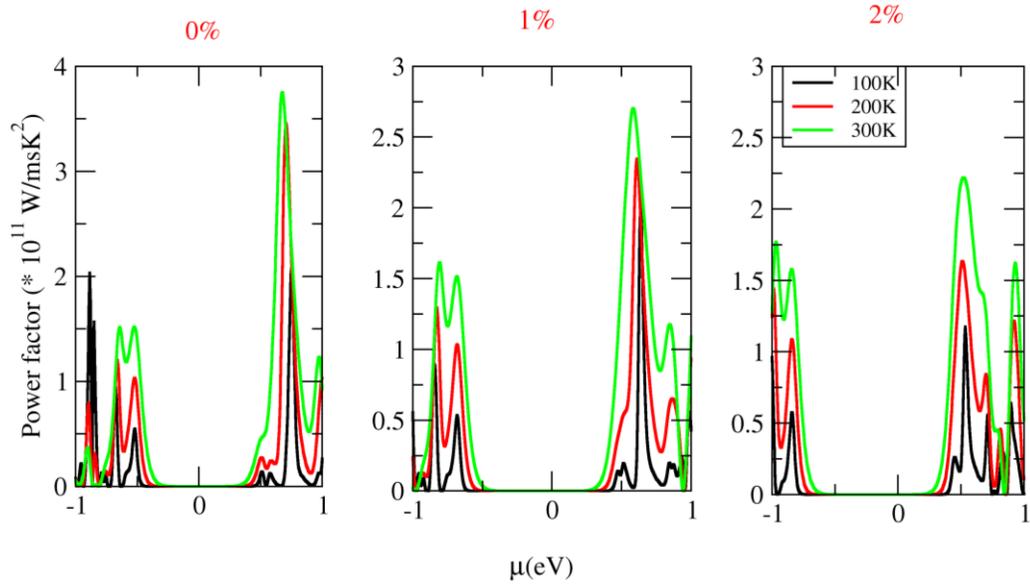

Figure 6: Power factor as a function of chemical potential at 100K, 200K and 300K.

In Figure 6 we have presented calculated power factor as a function of chemical potential under the application of strain at low temperatures i.e. 100K, 200K and 300K. We observed that for n-type doping PF is comparatively lower than p-type doping. For unstrained ISbTe, highest PF of $\sim 3.6 \times 10^{11} W/msK^2$ can be obtained at 300K with p-type doping. Although with strain application PF for n-type can be further increased from $\sim 1.5 \times 10^{11} W/msK^2$ to $2.6 \times 10^{11} W/msK^2$. Also, for p-type doping strain results in lowering the PF to $2.4 \times 10^{11} W/msK^2$ significantly due to comparatively higher thermal conductivity than unstrained system. Thus, strain has significant impact on enhancing band curvatures and hence the relevant effect can be observed in the power factor. Higher PF value represents the higher electronic contribution and is desirable for higher efficiency. Hence, strain application is a suitable strategy to modulate PF for optimizing parameters for higher ZT.

*3.3.5 Thermoelectric figure of merit (ZT)*

To predict efficiency of studied material, we have computed figure of merit for all the compositions as a function of chemical potential and as a function of temperature. Figure 7 depicts the computed ZT as a function of chemical potential at low temperatures. From the obtained values we can observe that for unstrained ISbTe Janus monolayer at 100K this material exhibit ZT ~0.98 with n-type and p-type dopings. At 300K, we can clearly observe that ZT increases to ~1.04 with p-type doping. Further applying strain on ISbTe by 1% shows increase in the ZT with value ~ 1.25 with p-type doping and with 2% strain this value increases to ~1.31 at 300K. This is significant due to optimized PF and lower thermal conductivity for p-type doping. Thus, our study primarily presents higher ZT with 2% strain at low temperature.

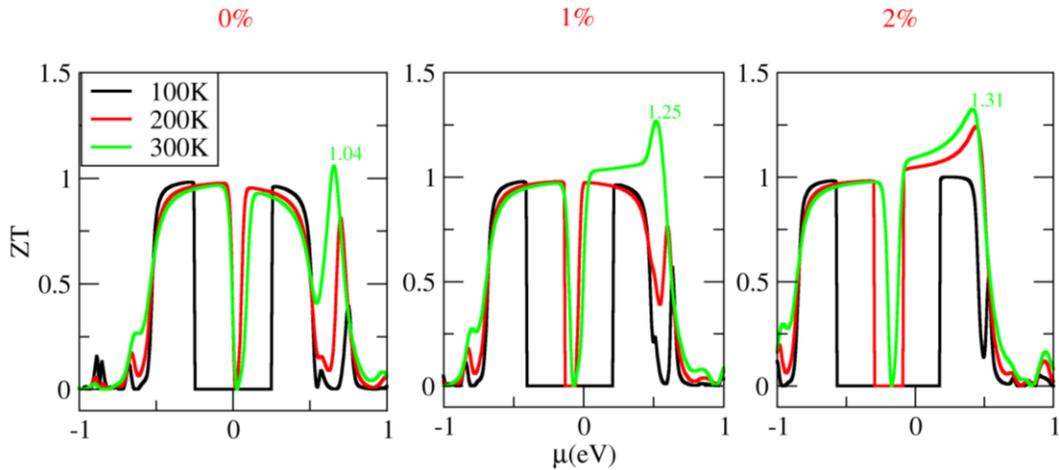

Figure 7: ZT as a function of chemical potential in different temperature regions.

To further investigate the variation of ZT with temperature, we have presented ZT as a function of temperature in Figure **8**. The results are shown for the constant chemical potential µ=0.42eV. From the figure 8, we can clearly see that at lower temperature say 250K we get the ZT values of ~0.81, ~0.89 and ~1.34 at 0%, 1% and 2% respectively. Whereas, at 300K we have obtained value of ZT as ~0.78, 1.04 and 1.34 for 0%, 1% and 2% strain. Thus, the studied Janus monolayer of ISbTe exhibits higher ZT values (>1) with strain which make them potential materials for thermoelectric applications.

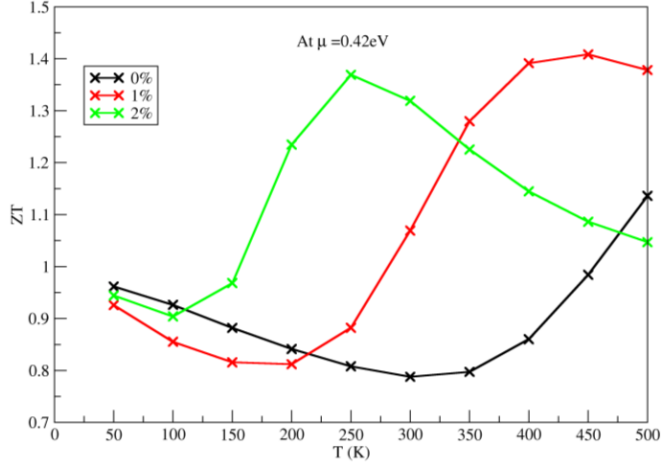

Figure 8: ZT as a function of temperature at constant chemical potential as 0.42 eV.

## Conclusions

The present study primarily focuses on the optimization of thermoelectric coefficients of Janus monolayer ISbTe in 2H phase by applying strain. We have systematically computed optimized structural parameters, dynamical stability of the material using first principles. Obtained structural parameters are in good agreement with the literature reports and phonon dispersion curves suggest that this material to withstand tensile strain up to 4%. Further, electronic band structures revealed this monolayer as indirect band gap semiconductor with band gap of 1.04eV which is comparable to previous reports as 1.21eV [68]. With rise in the strain by 2% band gap increases to 1.22eV. Thus, strain enhances the band curvatures and due to this significant change thermoelectric coefficients can be done. Computed thermoelectric coefficients suggests that this material has high Seebeck coefficient and lower electronic thermal conductivity which results in higher ZT >1. Our work implies that small strain of 2% with p-type doping can be a favorable strategy to obtain high ZT in the Janus monolayer of ISbTe at low temperatures.

## Acknowledgments

We would like to acknowledge the computational resources provided by the Indian statistical Institute, Kolkata, those have been helpful during this research. Also, we would like to acknowledge the developers of ELK code and Quantum espresso for providing free open-source codes.